\documentclass{sig-alternate}

\usepackage{amsfonts,amsmath,amssymb,latexsym,comment,enumerate}
\ifx\pdftexversion\undefined
\usepackage[a4paper,colorlinks,linkcolor=black,filecolor=black,citecolor=black,urlcolor=black,pdfstartview=FitH]{hyperref}
\else
\usepackage[a4paper,colorlinks,linkcolor=blue,filecolor=blue,citecolor=blue,urlcolor=blue,pdfstartview=FitH]{hyperref}
\fi
\usepackage{xspace}

\newtheorem{theorem}{Theorem}

\newtheorem{lemma}{Lemma}

\newtheorem{definition}{Definition}
\newtheorem{example}{Example}

\newcommand{\namedref}[2]{\hyperref[#2]{#1~\ref*{#2}}}
\newcommand{\sectionref}[1]{\namedref{Section}{#1}}

\newcommand{\lemmaref}[1]{\namedref{Lemma}{#1}}

\renewcommand{\P}{\mathcal{P}}

\newcommand{\ack}{\mathit{ack}}

\def\beginsmall#1{\vspace{-\parskip}\begin{#1}\itemsep-\parskip}
\def\endsmall#1{\end{#1}\vspace{-\parskip}}
\newenvironment{RETHM}[2]{\trivlist \item[\hskip 10pt\hskip\labelsep{\bf
#1\hskip 5pt\relax\ref{#2}.}]\it}{\endtrivlist}
\newcommand{\rethm}[1]{\begin{RETHM}{Theorem}{#1}}
\newcommand{\repro}[1]{\begin{RETHM}{Proposition}{#1}}
\newcommand{\relem}[1]{\begin{RETHM}{Lemma}{#1}}
\newcommand{\recor}[1]{\begin{RETHM}{Corollary}{#1}}
\newcommand{\erethm}{\end{RETHM}}
\newcommand{\erepro}{\end{RETHM}}
\newcommand{\erelem}{\end{RETHM}}
\newcommand{\erecor}{\end{RETHM}}

\newcommand{\commentout}[1]{}

\newcommand{\mwdavss}{\text{MW-SVSS}\xspace}
\newcommand{\davss}{{\text{SVSS}}\xspace}
\newcommand{\IN}{\mbox{$I\!\!N$}}
\renewcommand{\ell}{l}

\newcommand{\OK}{\mathit{OK}}

\newcommand{\A}{\text{ACK}\xspace}
\newcommand{\D}{\text{DEAL}\xspace}
\newcommand{\DMM}{\text{DMM}\xspace}

\begin{document}

\conferenceinfo{PODC'08,} {August 18--21, 2008, Toronto, Ontario, Canada.}

\CopyrightYear{2008}

\crdata{978-1-59593-989-0/08/08}

\title{An Almost-Surely Terminating Polynomial Protocol for
Asynchronous Byzantine Agreement with Optimal Resilience}

\numberofauthors{3}
\author{
\alignauthor Ittai Abraham\\
\affaddr{Hebrew University} \\
\email{ittaia@cs.huji.ac.il}
\alignauthor Danny Dolev\thanks{Part of the work was done
while the author visited Cornell university. The work was
funded in part by ISF, NSF, CCR, and AFOSR.} \\
\affaddr{Hebrew University} \\
\email{dolev@cs.huji.ac.il}
\alignauthor Joseph Y. Halpern\thanks{
Supported in part by NSF under grants ITR-0325453 and
IIS-0534064,  and by AFOSR under grant FA9550-05-1-0055.
} \\
\affaddr{Cornell University} \\
\email{halpern@cs.cornell.edu} }

\maketitle
\begin{abstract}
Consider an asynchronous system with private channels and $n$ processes,
up to $t$ of which may be faulty.  We settle a longstanding open
question by providing a Byzantine agreement protocol that simultaneously
achieves three properties:
\begin{enumerate}
\item \emph{(optimal) resilience}: it works as long as $n>3t$;
\item \emph{(almost-sure) termination}: with probability one, all
nonfaulty processes terminate;
\item \emph{(polynomial) efficiency}: the expected computation time,
  memory consumption, message size, and number of messages sent are all
  polynomial in $n$.
\end{enumerate}
Earlier protocols have achieved only two of these three
properties.  In particular, the protocol of Bracha is not polynomially
efficient, the protocol of Feldman and Micali is not optimally
resilient, and the protocol of Canetti and Rabin does not have
almost-sure termination. Our protocol utilizes a new primitive called
\emph{shunning (asynchronous) verifiable secret sharing}
(SVSS), which ensures,
roughly speaking, that either a secret is successfully shared or a new
faulty process is ignored from this point onwards
by some nonfaulty process.
\end{abstract}

F.0 {[Theory of Computation]}: {General}.


\section{Introduction}
The \emph{Byzantine agreement problem}, introduced in 1980 by Pease,
Shostak, and Lamport \citeyear{PSL80}, has emerged as one of the most
fundamental problems in Distributed Computing. The problem is easy to
describe:
each process has an input value; the goal is for
all processes to agree on a
consensus value that is an input value of one of the processes. The
challenge lies in reaching agreement despite the presence of faulty
processes. Many variants of the problem have been studied.  After three
decades of
extensive research, tight bounds have been obtained for almost all of
the variants, with one significant exception:
\emph{asynchronous Byzantine agreement}, where communication channels
between processes have unbounded delay
(although messages are guaranteed to arrive eventually),
and the faulty processes are
malicious in an arbitrary way
(though nonfaulty processes have secure private channels).

An execution of a Byzantine agreement protocol is
\emph{nonterminating} if some nonfaulty process 
does not output a value.
The celebrated result of Fischer, Lynch, and Paterson \citeyear{FLP}
shows that any protocol that never reaches disagreement
(i.e., has no executions where two nonfaulty processes output different
values)
must have
some nonterminating executions.
For a protocol that never reaches disagreement, the best we can hope for
is that the set of 
nonterminating executions has probability 0.
We say such protocols are  \emph{almost-surely terminating}.
Ben-Or \citeyear{Ben83} showed that almost-surely terminating asynchronous
Byzantine agreement can be achieved
as long as $n > 5t$,
where $n$ is the number of processes in the system and $t$ is a bound on
the number of faulty processes.  However, his protocol required an
expected number of rounds that is exponential in $n$.
This started a lengthy sequence of research on asynchronous Byzantine
agreement; see, for example,
\cite{Ben83,Bra84,CR93,FM88,KK06,RB89}.
It is well known that Byzantine agreement for $n$ processes cannot be
reached if $n\leq 3t$ \cite{PSL80}.
Therefore the best resilience one can hope for is $n>3t$. We
will call such protocols \emph{optimally resilient}.
Bracha \citeyear{Bra84}
provides an almost-surely terminating protocol that is optimally
resilient.
However his
protocol does not scale well with the size of the system, since,
like Ben-Or's, the
expected number of messages and rounds is exponential in $n$.
Feldman and Micali \citeyear{FM88}
provide a Byzantine agreement protocol for the synchronous model with
optimal resilience and constant expected running time.
They extend their result to the asynchronous model,
where they provide a
polynomial-time algorithm that almost-surely terminates, but
does not have optimal resilience; their protocol requires that $n > 4t$.
Canetti and Rabin \cite{CR93,CR93-long} provide a protocol that is
optimally resilient $(n>3t)$ and polynomially efficient.
Their result uses ideas from Rabin and Ben-Or \citeyear{RB89} on
verifiable secret sharing (VSS) in synchronous systems equipped with a
broadcast channel. The techniques of \cite{RB89} have an inherent
nonzero probability of failure; as a result, in the asynchronous
implementation of \cite{CR93},
the protocol is not  almost-surely terminating.
Indeed, in \cite{CR93-long}, the authors explicitly highlighted the problem
of finding a protocol that
simultaneously achieves optimal resilience, almost-sure termination, and
polynomial efficiency.
Up to now, despite repeated efforts, this has not been done.
The main result of this paper is to provide such a protocol.

Pretty much all protocols following Bracha's \citeyear{Bra84} used his
idea of reducing the problem of Byzantine agreement to that of
implementing a shared coin. We do that as well.
We obtain a shared coin using an approach that goes back to Feldman
and Micali \citeyear{FM88,FM97}, who essentially reduce
the problem of efficiently implementing a shared coin to that of
efficiently implementing VSS.
Roughly speaking, the secrets in VSS are used to generate a shared coin.
We refer
the reader to Canetti's thesis \cite{Can-thesis} (Chapters 4 and 5)
for a comprehensive account of the rather complex reduction from
VSS to Byzantine agreement in the asynchronous model.

The protocol of Canetti and Rabin~\cite{CR93}
also
uses the reduction from
verifiable secret sharing to Byzantine agreement.
The only reason that their protocol is not
almost-surely terminating is that they use a protocol that they call
Asynchronous Verifiable
Secret Sharing (AVSS), which has a small 
(but nonzero)
probability of not terminating.
Our protocol has essentially the same structure as the Canetti-Rabin
protocol.
Indeed, it
uses the same reduction from AVSS to Byzantine agreement as
in  \cite{CR93},
except that the use of AVSS is replaced by
a variant of AVSS that we call
\emph{shunning (asynchronous) verifiable secret sharing} (\davss).
which is guaranteed to terminate almost-surely.

To explain
the properties of \davss, we first review
the properties of standard
VSS
(verifiable secret sharing).
VSS involves a dealer who has a value to share, which we think of as the dealer's
secret.
It has two key
properties, known as \emph{validity} and \emph{binding}.
Informally, the validity property
guarantees that, if the dealer is nonfaulty, then all nonfaulty processes will
reconstruct
the
dealer's value; the binding property guarantees that a faulty
dealer must commit to a value during what is called the \emph{share
phase} of the protocol.
Our \davss scheme has weaker validity and binding properties.
Specifically, we require that in each invocation where the validity or binding
properties do not hold,
at least one nonfaulty process
ignores
at least one new faulty
process
from that invocation on.
The key observation is that this limits the
adversary to breaking the validity and binding properties at most a
polynomial number of times.

The \davss protocol uses a weaker protocol called \emph{moderated weak
shunning (asynchronous) VSS} (\mwdavss).
The
\mwdavss protocol is a variant of VSS
with a dealer and an additional entity called a
\emph{moderator}.
In \mwdavss the dealer has some input value $s$ and the moderator has some input value $s'$.
The nonfaluty moderator's task is to enforce during the share phase that the value that the dealer shares is $s'$ (hence $s=s'$ if both are nonfaulty).
The initials MWS characterize how \mwdavss differs from standard VSS:
\beginsmall{itemize}
\item \textbf{Moderated}. A  faulty dealer must commit
to
the value of the nonfaulty moderator in order to complete the share
protocol.
(Katz and Koo \citeyear{KK06} use a moderator for VSS in a
somewhat similar way.)
\item \textbf{Weak}. As in \emph{weak VSS} \cite{CR93,RB89}, the binding
property of VSS is weakened so that each process
reconstructs either
the committed value or a default value (denoted $\bot$).
\item \textbf{Shunning}. Like \davss,
it is possible that neither
validity nor the weaker binding property hold, but in that case
at least one nonfaulty process
ignores
at least one new faulty process
from this stage on.
\endsmall{itemize}

As in the VSS scheme used in \cite{BGW88,FM88,FM97}, the \davss scheme
starts with a dealer who shares
a degree-$t$ bivariate polynomial
$f(x,y)$ such that $f(0,0)$ is the secret.
Each process $i$ gets $t+1$ values of each of the
polynomials $g(y) = f(i,y)$ and $h(x) = f(x,i)$,
which is enough to reconstruct them, since they both have degree $t$.
Then, roughly speaking,
each pair $(i,j)$ of processes uses \mwdavss to commit to
$f(i,j)$ and $f(j,i)$.
This ensures that if either $i$ or $j$ is nonfaulty then the
reconstructed values of the \mwdavss protocol will be either $\bot$ or
the required values ($f(i,j)$, $f(j,i)$). We then use this fact to
prove the properties of the \davss protocol.
The key property of the \mwdavss protocol is its
use of a fault-detection mechanism. The mechanism has the property that
a nonfaulty process might not explicitly know it has detected a
faulty process.
The only guarantee is that it will act as if it has detected a faulty
process, by ignoring all messages from the detected process for the rest
of the protocol.
This behavior is somewhat reminiscent of the failure detector $\diamond
\mathcal{W}$ \cite{CHT96} in the sense that a nonfaulty process might
reach a state
of permanently suspecting a faulty process without being explicitly
aware of this fact.
Since the details of the \mwdavss protocol are somewhat technical, we refer the reader to \sectionref{sec:high-level}
for a high-level description.

The rest of this paper is organized as follows.
In Section~\ref{sec:properties}, we state the properties of \davss
and \mwdavss.  In Section~\ref{sec:MWFD-AVSS}, we  provide an
implementation of \mwdavss and prove that it has the required
properties.  In Section~\ref{sec:FD-AVSS}, we do the same for \davss,
using \mwdavss as a subroutine.
A description of Bracha's
\emph{Reliable Broadcast} protocol, which we use as a subroutine, is
given in the appendix.

\section{Shunning VSS}\label{sec:properties}
As we mentioned above,
in \davss, if
either
the binding property or the validity property
does not hold, then a new faulty process is
ignored in all future invocations
by some nonfaulty
process.
To implement this, each process needs to keep track of the processes it
knows to be faulty.  Thus, the \davss scheme actually has two
components: a \emph{detection and message management protocol} (\DMM
protocol) and a \emph{VSS protocol}.  Each process uses its \DMM protocol
to decide which messages to discard, which to ignore for now, and which
to act on, and to keep track of the processes it knows to be faulty.
The \DMM protocol is invoked when the \davss scheme is initialized, and
then runs indefinitely and concurrently with all the invocations of the
VSS protocols.
The VSS protocol may be invoked a number of times while the \davss scheme runs, and several invocations may be running concurrently.
The VSS protocol is composed of a pair of protocols $\mathcal{S}$ (for
\emph{share}) and $\mathcal{R}$ (for \emph{reconstruct}).
These protocols are called separately; $\mathcal{R}$ is never called
unless $\mathcal{S}$ completes, but $\mathcal{R}$ may not be called at
all even if $\mathcal{S}$ completes.
We associate with each VSS invocation a unique session identifier
$(c,i)$ that is composed of
a counter $c$ and
the dealer's identifier $i$.
We tag all events of that invocation with its session identifier,
so that it is always clear which invocation of the VSS protocol an event
belongs to.

We say that a VSS invocation has \emph{completed for process $j$} if process
$j$ completed the reconstruct associated with that
session. Given a process $j$ and two VSS invocations with session
identifiers $(c,i)$ and $(c',i')$, we write $(c,i) \rightarrow_j
(c',i')$ if process $j$ completes the invocation of the VSS $(c,i)$
before process $j$ begins the invocation of the VSS $(c',i')$.

As we said in the introduction, our VSS scheme is \emph{shunning}.
Process $i$ may start shunning $j$ well before $i$ is sure that
$j$ is faulty;
indeed, $i$ may shun $j$ without ever knowing that $j$ is faulty.
\begin{definition}
Process $j$ is \emph{shunned} by process $i$ starting in session
$(c,\ell)$
of \mwdavss (resp., \davss)
if
process $i$ does not ignore some message from $j$ during
session $(c,\ell)$,
but
ignores or discards
all messages from $j$ associated with
every
session $(c',\ell')$
of  \mwdavss (resp., \davss)
such that $(c,\ell) \rightarrow_i (c',\ell')$.
\end{definition}
\subsection{Properties of \davss}
Each VSS invocation has one process $d$ designated as the
\emph{dealer}; the
dealer has some input value $s$.
For ease of exposition, we do not include the session identifier in our
description of the properties of the VSS protocol when they are clear
from the context, although we do include them in the description of the
protocols.
Each VSS invocation must satisfy the following properties (in runs with at
most $t$ faulty processes); we call these the
\emph{\davss} properties.
\beginsmall{enumerate}
\item \textbf{Validity of Termination}. If a nonfaulty dealer initiates
    protocol $\mathcal{S}$, then each nonfaulty process will eventually
    complete protocol $\mathcal{S}$.

\item \textbf{Termination}. If a nonfaulty process completes protocol
    $\mathcal{S}$, then all nonfaulty processes will eventually complete
    protocol $\mathcal{S}$.
    Moreover, if all nonfaulty processes begin protocol $\mathcal{R}$,
    then all nonfaulty processes will eventually complete
    protocol $\mathcal{R}$
(note, however, that if only some but not all nonfaulty processes begin
protocol $\mathcal{R}$, then there is no termination requirement).
\item \textbf{Binding}. Once the first nonfaulty process completes an invocation of
$\mathcal{S}$ with session id $(c,d)$,
    there is a value $r$ such that either
 \beginsmall{itemize}
 \item the output of each nonfaulty process that completes protocol
 $\mathcal{R}$ is $r$; or
 \item  there exists a nonfaulty process $i$ and a faulty process $j$ such that $j$ is shunned by $i$ starting in session $(c,d)$.
\endsmall{itemize}

\item  \textbf{Validity}. If the dealer is nonfaulty, then either
     \beginsmall{itemize}
         \item  the output of each nonfaulty process that completes protocol $\mathcal{R}$ is $s$; or
         \item  there exists a nonfaulty process $i$ and a faulty process $j$ such that $j$ is shunned by $i$ starting in session $(c,d)$.
    \endsmall{itemize}

\item \textbf{Hiding}.  If the dealer is nonfaulty and no
    nonfaulty process invokes protocol $\mathcal{R}$, then the
    faulty processes learn nothing about the dealer's value%
    \footnote{To make this precise, assume that the adversary
    determines the \emph{scheduling protocol}:
    how long each message will take to arrive as a function
    of the history.  Note that once we fix the inputs, the faulty processes,
    the protocols used by the faulty processes, and the scheduling
    protocol, the VSS protocol (which is used by the nonfaulty processes)
    determines a distribution on runs.  Formally, hiding requires that for
    all distributions
    determined this way, the dealer's value is independent of the
    histories of the faulty processes.}
\endsmall{enumerate}

\subsection{Properties of \mwdavss}
In order to implement the VSS protocol, we use a weaker protocol called
\emph{moderated weak shunning (asynchronous) VSS} (\mwdavss).
Just as VSS, the \mwdavss protocol is composed of a share protocol
$\mathcal{S}'$
and a reconstruction protocol $\mathcal{R}'$.
As in weak VSS, we weaken the Binding property so that each
nonfaulty process reconstructs either $r$ or $\bot$.
But now, in addition to
having one process $d$ designated as the dealer,
there is
an additional
process designated as the
\emph{moderator}.
Both the dealer and the moderator have (possibly different) input
values, denoted $s$ and $s'$,
respectively.
Each \mwdavss invocation must satisfy Termination and Validity,
just like VSS, and the following variants of the properties of VSS
(in runs with at most $t$ faulty processes); we call these the
\emph{\mwdavss} properties.
\beginsmall{enumerate}
\item [1$'$.] \textbf{Moderated Validity of Termination}. If a nonfaulty
    dealer initiates protocol $\mathcal{S}'$, the moderator is nonfaulty, and $s=s'$,
    then each nonfaulty process will eventually complete protocol $\mathcal{S}'$.

\item [3$'$.] \textbf{Weak and Moderated Binding}.
Once the first nonfaulty process completes
an invocation of protocol $\mathcal{S}'$ with session id $(c,d)$,
there is a value $r$ (possibly $\bot$) such that
 \beginsmall{itemize}
 \item if the moderator is nonfaulty, then $r=s'$.
 \endsmall{itemize}
In addition, either
 \beginsmall{itemize}
 \item the output of each nonfaulty process that completes protocol
    $\mathcal{R}'$ is either $r$ or $\bot$; or
 \item  there exists a nonfaulty process $i$ and a faulty process $j$
    such that $j$ is shunned by $i$ starting in session $(c,d)$.
 \endsmall{itemize}

\item [5$'$.] \textbf{Moderated Hiding}.  If the dealer and moderator
are nonfaulty and no nonfaulty process invokes protocol
$\mathcal{R}'$, then the faulty processes learn nothing about
the dealer's value.
\endsmall{enumerate}

It might seem surprising that in the second condition of Validity and
(Weak and Moderated) Binding, we talk about shunning rather than just
saying that a faulty process is detected.  The reason is that, as we
show in Example~\ref{xam:shunning} (after we give the implementation of
the \mwdavss protocol), it is possible that two nonfaulty processes
will complete an invocation of the \mwdavss protocol
with different values
without
(explicitly)
detecting
a new faulty process; however, in that case, at least one of them will
shun a faulty process that was not shunned before.

\section{Implementing \DMM and \mwdavss}\label{sec:MWFD-AVSS}
\subsection{A high-level description}\label{sec:high-level}
In this section, we provide an implementation of \DMM and \mwdavss.
We start with
a high-level description of both.
Both protocols use the Reliable Broadcast protocol (RB) of
Bracha \cite{Bra84}.
RB guarantees that messages are
indeed broadcast; if a nonfaulty sender sends a message $m$, then all
nonfaulty processes eventually receive $m$, and nothing else.
(The properties of RB are stated carefully in the appendix,
where, for completeness, an implementation is provided.)

We assume that the dealer
creates $n+1$ degree-$t$ polynomials $f,f_1,\dots,f_n$
over some finite field $F$ with $|F| > n$
such that $f(0)$
is the secret
(i.e., $f(0) = s$)
and $f_\ell(0)=f(\ell)$. Then the dealer shares the
polynomials $f_1,\dots,f_n$  and also gives
each process $j$ the polynomial $f_j$.
We can think of process $j$ as a
potential ``monitor'' for $f_j$.
The dealer shares the polynomial $f_j$ by sending each process $k$ the
value $f_j(k)$.  This means that, if the dealer is correct, any $t+1$
nonfaulty processes can reconstruct $f_j$.
In addition, the dealer sends $f$ to the moderator.
Each process $k$ that receives $f_j(k)$ sends this value to $j$ and
broadcasts a confirmation.
In this case, we can think of process $k$ as a ``confirmer'' for $f_j(k)$.
When
$j$ receives confirmations and values that agree with the polynomial
$f_j$ sent by the dealer from at least $n-t$ processes,
$j$ becomes a ``monitor'' for $f_j$, sends
$f_j(0)$ to the moderator, and broadcasts the set $L_j$ of
at least
$n-t$ confirmers
whose value it
accepted. Intuitively, each monitor $j$
is responsible for validating
the value of one
point on the polynomial $f$, namely,
$f(j)=f_j(0)$.
When the moderator receives
at least
$n-t$ values all of which agree with the polynomial $f$ from different
monitors
and receives confirmations from their associated $L_j$ sets, then the moderator
broadcasts the
set of $n-t$ monitors' indexes it accepted. The dealer broadcasts a
confirmation when it learns that the moderator, its monitors, and their
confirmers have
acted in a nonfaulty manner.
This allows nonfaulty processes to know which confirmers they need to
wait for in order to complete their execution of the share protocol.
In the reconstruct phase, processes send their values using the RB
protocol. If the dealer is nonfaulty, then
it can check  the values sent by all processes and detect faulty processes.
If the dealer is faulty, then there are at least $t+1$ nonfaulty monitors
$\ell$
that can monitor their polynomial
$f_\ell$.
If they do not detect
problems
with their confirmers,
then the
Weak Binding property must hold.

We now explain how processes shun other processes if a problem is detected.
Before a process $i$ ``sees'' a message in the \mwdavss protocol (or the
\davss protocol that we present later),
the message is filtered by the DMM protocol.
The $\DMM_i$ protocol
decides whether to discard the message, ignore it for now, or pass it on
for action.
In order to do this, $\DMM_i$ must maintain a number of data
structures.
First, it maintains the partial order $\rightarrow_i$ on
sessions described above,
where $(c_1,j_1) \rightarrow_i (c_2,j_2)$ if $i$ started the
share protocol of VSS session $(c_2,j_2)$ after completing the
reconstruct protocol of VSS session $(c_1,j_1)$.
In addition,
the $\DMM_i$ protocol uses a variable $D_i$ that
represents a set of processes.
Intuitively, the processes in $D_i$ are ones known by $i$ to be faulty.
Any message sent by a process $j \in D_i$ is discarded by $i$.
To decide which messages to ignore for now and which to pass on for
action, $\DMM_i$ maintains two arrays.
The first array, denoted $\A_i$, consists of tuples in
$\{1, \ldots, n\} \times \{1, \ldots, n\} \times \IN \times F$.
Intuitively, $(j,l,c,x) \in \A_i$ if $i$ is expecting to receive a
broadcast sent by $j$ using RB saying $f_l(j) = x$ as part of a
VSS session $(c,i)$ (thus, this is a session for which $i$ is the dealer).
The second array, denoted $\D_i$, consists of tuples in
$\{1, \ldots, n\} \times \IN \times
\{1,\ldots, n\} \times F$.  Intuitively,
$(j,c,\ell,x) \in \D_i$ if $i$ is expecting to receive a
message broadcast by $j$ (using RB) saying $f_i(j) = x$
as part of VSS session $(c,l)$.
Both $\A_i$ and $\D_i$ are initially empty.
We will explain how tuples are added to $\A_i$ and $\D_i$ when
we describe the \mwdavss protocol.

Process $i$ ignores
(that is, saves but does not act on)
all messages from process $j$ that are part of a
session $(c',k)$  such that either $(j,l,c,s) \in \A_i$ and $(c,i)
\rightarrow_i (c',k)$ or
$(j,c,l,s) \in \D_i$ and $(c,l) \rightarrow
(c',k)$.  That is,
newer
messages from $j$ are ignored by $i$ if $i$ is
expecting to receive something from $j$ that it has not yet received.
When a message that $i$ expects to hear from $j$ that is associated with
either with  $(j,l,c,s) \in \A_i$ or with $(j,c,l,s) \in \D_i$, then the
relevant tuple is removed from $\A_i$ or
$\D_i$.
Once there are no messages that $i$ expects to hear from $j$
from a session that precedes $(c',k)$, then the $\DMM_i$ protocol enables
the \mwdavss protocol to act on messages from session $(c',k)$.

Finally, process $j$ is added to $D_i$ if a message is received from $j$
that is inconsistent with what is expected according to a tuple in
$\A_i$ or $\D_i$.  For example, if
$(j,l,c,s) \in \A_i$ and $i$ receives a message as part of session
$(c,i)$ from $j$ saying $f_l(j) = s'$, with $s' \ne s$, then $j$ is
added to $D_i$,
and messages sent by $j$ in all sessions $(c',k)$ such that $(c,l)
\rightarrow_i (c',k)$ will be
discarded by $i$.

\subsection{Implementing \mwdavss}
We now show how to implement \mwdavss.  We start with the
share protocol $\cal{S}'$.  We assume that the field $F$ being used is
common knowledge and $|F| > n$.
In the $\mathcal{S}'$ protocol (and a number of our later protocols), we
have variables that are tagged by the session id $(c,d)$.  If the
session id is clear from context, we omit it.

\subsubsection*{Share protocol $\mathcal{S}'$:}
\begin{enumerate}
\item If a dealer $i$ wants to invoke $\mathcal{S}'$ with a
secret $s$ it first updates $c$ to $c+1$ and then
selects $n+1$ random degree-$t$
polynomials $f(x),f_1(x),\dots,f_n(x)$ over field $F$ such
that $f(0)=s$ and $f_\ell(0)=f(\ell)$ for all
$\ell \in \{1, \dots, n\}$.
It sends each process $j$ a message
 $f_1(j),\dots,f_n(j),(c,i)$.
In addition, it sends each process $\ell$ a message
$f_\ell(1),\ldots, f_\ell(t+1),(c,i)$ (note that this allows $\ell$ to
compute $f_\ell$, so we sometimes say ``$\ell$ receives $f_\ell$'' in
this message),
and sends the moderator yet another message,
$f(1), \ldots, f(t+1),(c,i)$ (so that the moderator can compute $f$).

\item If process $j$ receives values $\hat{f}^j_1,\dots, \hat{f}^j_n$
and polynomial $\hat{f}_j$ from a dealer $i$ in session $(c,i)$,
then, for each process $\ell$, $j$ sends
$\hat{f}^j_\ell,(c,i)$ to $\ell$.
(Note that $\hat{f}^j_k$ is supposed to be $f_k(j)$, but if the dealer
is faulty, it may not be.  We continue to use the notation $\hat{f}$ and
$\hat{f}^j_k$ to denote the polynomials and values actually received.)
It also broadcasts $\ack,(c,i)$ to all processes using RB.

\item If process $j$ receives $\hat{f}^\ell_j,(c,i)$  and
$\ack,(c,i)$ from process $\ell$, receives $\hat{f}_j,(c,i)$
from the dealer $i$, and $\hat{f}^\ell_j=\hat{f}_j(\ell)$, it
adds $(\ell,c,i,\hat{f}_j(\ell))$ to $\D_j$.
Intuitively, the message $\hat{f}^\ell_j=\hat{f}_j(\ell)$ provides
confirmation to $j$ that
the dealer sent $f_\ell(j)$ to both $j$ and $\ell$.  The fact that
$j$ adds $(\ell,c,i,\hat{f}^\ell_j)$ to $\D_j$ means that $j$
expects $\ell$ to confirm publicly (using RB) that indeed it received
$\hat{f}^\ell_j$ from the dealer $i$, which is what $\ell$ told $j$
privately.

\item Let $L_{j,(c,i)} = \{\ell: (\ell,c,i,\hat{f}_j(\ell)) \in \D_j\}$.
If $|L_j| \ge n-t$,
then  $j$ sends $L_j,(c,i)$ to all processes using RB,
It also sends $\hat{f}_j(0),(c,i)$ to the moderator.  Intuitively, if $|L_j|
\ge n-t$, then $j$ has gotten as much confirmation as it can expect
to get that the dealer $i$ correctly shared the polynomial
$f_j$.  By broadcasting $L_j$, it is broadcasting the set of processes
from which it
expects to hear public  confirmation of this fact.  By sending
$\hat{f}_j(0)$ to the moderator, $j$ is giving the moderator a
share of the information that the moderator needs
for computing the secret.
\item If the moderator receives $\hat{f},(c,i)$ from the dealer,
$\hat{f}^j_0,(c,i)$ and $\hat{L}_j,(c,i)$ from process $j$, and
$\ack,(c,i)$ message from all processes $\ell \in \hat{L}_j$,
$\hat{f}^j_0=\hat{f}(j)$, and $\hat{f}(0)=s'$, the moderator
adds $j$ to the set $M_{(c,i)}$,
which is initialized to $\emptyset$.
Intuitively, if the values that the
moderator receives from $j$ are compatible with the values the
moderator received from the dealer, and the dealer's values are
compatible with the moderator's value $s'$,
then the moderator adds $j$ for the session $(c,i)$ to $M$.
\item
If
$|M_{(c,i)}| \ge n-t$, the
moderator sends $M_{(c,i)},(c,i)$ to all processes using RB.

\item If the dealer $i$ receives $\hat{M},(c,i)$ from the moderator,
receives $\hat{L}_j,(c,i)$ from  each process $j \in \hat{M}$,
and receives $\ack,(c,i)$ from each process $\ell \in \hat{L}_j$ such
that $j \in  \hat{M}$, then
it adds  $(\ell,j,c, f_j(\ell))$ to $\A_i$
for all $j
\in \hat{M}$ and $\ell\in \hat{L}_j$, and sends $\OK,(c,i)$
using RB.
Note that if the moderator is nonfaulty and it sends these messages to
the dealer, then it really did receive
$\hat{L}_j,(c,i)$ from each process $j \in \hat{M}$
and $\ack,(c,i)$ from each process $\ell$ in $\hat{L}_j$, and these
messages were sent
using RB.  Thus, the dealer will eventually receive all these
messages too and, if nonfaulty, will broadcast the $\OK$ message.

\item If process $j$ receives $\hat{M},(c,i)$ from the moderator and $j
\notin \hat{M}$ then 
$j$ removes from $\D_j$ all entries of the form $(\cdot,c,i,\cdot)$ that
are associated with session $(c,i)$. Intuitively, since 
$j \notin \hat{M}$ for session $(c,i)$, we do not care about the values
of $f_j$ for this session.

\item If process $j$ receives an $\OK,(c,i)$ message from the dealer,
$\hat{M},(c,i)$ from the moderator, $\hat{L}_\ell,(c,i)$ from each
process $\ell
\in \hat{M}$, and $\ack,(c,i)$ from each $k \in \hat{L}_\ell$ such that
$\ell \in
\hat{M}$, it completes
this invocation of the share protocol $\mathcal{S}'$.

\end{enumerate}

\subsubsection*{Reconstruct protocol $\mathcal{R}'$:}

\begin{enumerate}
\item
If process $j \in 
\hat{L}_\ell
$ for $\ell \in \hat{M}$, then $j$ broadcasts
$\ell, \hat{f}^j_\ell,  (c,i)$
using RB, where $\hat{f}^j_\ell$ is what $j$ received from the dealer at
step 2 of $\mathcal{S}'$.

\item
Process $j$ initializes $K_{j,\ell,(c,i)}$ to  $\emptyset$ for each process
$\ell$ for which it has received a set $L_\ell$.
If $j$ receives a message $\ell,\bar{f}^k_\ell,(c,i)$ from
process $k$ at step 1, and  $k \in \hat{L}_\ell$, then $j$ adds
$(\ell,\bar{f}^k_\ell)$
to $K_{j,\ell}$.
Intuitively, $(\ell,\bar{f}^k_\ell)$ should be the point
$(k,f_\ell(k))$ on the polynomial $f_k$.

\item
If $|K_{j,\ell}| = t+1$, then $j$
finds the unique degree $t$ polynomial
$\bar{f}_\ell$ that interpolates the points in $|K_{j,\ell}|$.

\item After computing $\bar{f}_\ell$ for all $\ell \in\hat{M}$, $j$
tries to interpolate a polynomial $\bar{f}$
such that $\bar{f}(\ell)=\bar{f}_\ell(0)$ for all $\ell \in \hat{M}$.
If $\bar{f}$ exists, $j$ outputs $\bar{f}(0)$;
otherwise, $j$ outputs $\bot$.
\end{enumerate}

\subsection{Implementing \DMM}
We now describe the implementation of $\DMM_i$.

\medskip
\textbf{Protocol $\DMM_i$}
\vspace{-2mm}
\beginsmall{enumerate}

\item Initialize an empty set of processes $D_i$, an empty array $\A_i$
consisting of tuples in $\{1, \ldots, n\} \times \{1, \ldots, n\} \times
\IN \times F$, and an empty array $\D_i$ consisting of tuples in $\{1,
\ldots, n\} \times \IN \times \{1,\ldots, n\} \times F$.
As we said earlier,
intuitively, $(j,l,c,x) \in \A_i$ if $i$ is expecting to receive a
    broadcast sent by $j$ using RB saying $f_l(j) = x$ as part of VSS
session $(c,i)$
and
    $(j,c,\ell,x) \in \D_i$ if $i$ is expecting to receive a
    message broadcast by $j$ using RB saying $f_i(j) = x$
    as part of VSS session $(c,l)$.

\item If $(j,\ell,c,x) \in \A_i$ and a broadcast message $x',j,(c,i)$ is
received then
    \beginsmall{itemize}
    \item if $x = x'$, then remove $(j,\ell,c,x)$ from $\A_i$;

    \item otherwise, add $j$ to $D_i$.

    \endsmall{itemize}

    (See line 7 of protocol $\mathcal{S}'$ for the condition that causes
    a tuple $(j,\ell,c,x) $ to be added to $\A_i$.)

\item   If $(j,c,\ell,x) \in \D_i$ and a broadcast message $x',i,(c,j)$
is received, then
    \beginsmall{itemize}
    \item if $x = x'$ then remove $(j,c,\ell,x)$ from $\D_i$;

    \item otherwise, add $j$ to $D_i$.
    \endsmall{itemize}

    (See line 3 of protocol $\mathcal{S}'$ for the condition that causes
    a tuple $(j,c,\ell,x) $ to be added to $\D_i$.)

\item If a message sent from $j$ is received and $j \in D_i$, then
discard the message.

\item If a message with session identifier $(c',i')$ sent from $j \notin
D_i$ is received, then delay this message if there is a tuple
$(j,\ell,c,x) \in \A_i$ such that $(c,i) \rightarrow_i (c',i')$ or a
tuple  $(j,c,\ell,x) \in \D_i$ such that $(c,j) \rightarrow_i (c',i')$.
If there is no such tuple in $\A_i$ or $\D_i$ (or after all such tuples
have been removed), then forward the message to the
   VSS invocation of session $(c',i')$.

\endsmall{enumerate}

We now show that the \mwdavss protocol satisfies the \mwdavss
properties.  To do this, we first must establish two key properties of
the \DMM protocol.

\begin{lemma}\label{lem:DMM}
If $i$ is nonfaulty, then $\DMM_i$ satisfies the following
two properties:
\beginsmall{itemize}
\item[(a)]  if $j \in D_i$, then $j$ is a faulty process;
\item[(b)] if $j$ is nonfaulty,
$(j,l,c,x) \in \A_i$ (resp., $(j,c,\ell,x) \in \D_i$),
and all 
nonfaulty
processes complete session $(c,i)$ (resp. $(c,\ell)$), then
eventually
$(j,l,c,x)$ is removed from $\A_i$ (resp., $(j,c,\ell,x)$ is removed
from $\D_i$).
\endsmall{itemize}
\end{lemma}

\begin{proof}
For part (a), note that
the only reason that $i$ adds $j$ to $D_i$ is if $(j,\ell,c,x) \in \A_i$
(resp., $(j,c,\ell,x) \in \D_i$) and the \DMM protocol detects that
process $j$ sent a message $\bar{f}_\ell^j, \ell, (c,i)$
(resp., $\bar{f}_i^j, i,(c,\ell)$)
using RB
such that $\bar{f}_\ell^j \neq x$
(resp., $\bar{f}_i^j \neq x$). If $j$ is nonfaulty then $x=f_\ell(j)$
(resp., $x=\hat{f}_i(i)$),
hence $i$ would not add $j$ to $D_i$ if $j$ is nonfaulty.

Part (b) follows from the observation that if $(j,\ell,c,x)
\in\A_i$ or $(j,c,\ell,x) \in \D_i$, then the tuple was added during
the share phase.
If $(j,\ell,c,x) \in\A_i$ and session $(c,i)$ completed, then it must
be the case that $j \in \hat{L}_\ell$ and $\ell \in
\hat{M}_{(c,i)}$. Since $j$ is nonfaulty, then the message required to
remove the tuple from $\A_i$ will be sent 
using RB by $j$ during the reconstruct phase, and will eventually be
received by $i$. 
If $(j,c,\ell,x) \in \D_i$ then there are two cases. If this entry was
removed in line 8 of protocol $\mathcal{S}'$, then we are
done. Otherwise, since session $(c,\ell)$ completed, it must be the
case that $j\in \hat{L}_i$ and $i \in M_{(c,\ell)}$. Hence the message
required to remove the tuple from $\D_i$ will be sent using RB by $j$
during the reconstruct phase, and will eventually be received by $i$. 
\end{proof}

We now prove that all the \mwdavss properties hold.
\begin{lemma}\label{lem:mwdavss}
The \mwdavss protocol satisfies the\\ \mwdavss properties.
\end{lemma}
\begin{proof}
We consider the properties in turn.

\textbf{Moderated Validity of Termination}. If the dealer and the
moderator are nonfaulty and $s=s'$ then, for
all nonfaulty processes $j$ and $\ell$, eventually $(j,c,i,\hat{f}^j_l)$
will be in $\D_\ell$.   Hence, eventually $|L_\ell|$ will be at least
$n-t$.  Thus, eventually $\ell$ will complete step 4 of the share protocol.
(For future reference, note that although the first $n-t$ elements of
$L_\ell$ may not all be nonfaulty,
at least $t+1$ of the elements of $L_\ell$ will be nonfaulty.)
Moreover, since $j' \in L_\ell$ only if $j'$ sent an $\ack,(c,i)$ message
using RB, eventually the moderator will receive an
$\ack,(c,i)$ message from all $j \in L_\ell$.  Thus, if $\ell$ is
nonfaulty, a nonfaulty moderator will eventually add $\ell$ to $M$
in step 5 of the share protocol.
Since there are $n-t$ nonfaulty processes,
eventually we must have
$|M| \ge n-t$, so the moderator completes step 6 of the share
protocol.  We already gave the intuition that a nonfaulty dealer will then
broadcast $\OK$ at step 7.  Thus, all nonfaulty processes
will eventually complete protocol $\mathcal{S}'$.

\textbf{Termination}. If a nonfaulty process $j$ completes protocol
$\mathcal{S}'$,
then, since all the messages that caused $j$ to complete the protocol
are sent using RB, it follows that all nonfaulty processes
eventually complete $\mathcal{S}'$.
The fact that
they all  complete $\mathcal{R}'$ follows since, as observed above, the
set $L_\ell$ for each
$\ell \in M$ contains at least $t+1$ nonfaulty
processes, each of which eventually sends its value
in step 1 of $\mathcal{R}'$.  Thus, each nonfaulty process
outputs either some value in $F$ or $\bot$ at step 3 of $\mathcal{R}'$.

\textbf{Validity}.
Suppose that the dealer $i$ is nonfaulty.  There are two cases.   If
some  faulty process $j$ such that $(j, \ell, c, x) \in \A_i$ sends
a message $x',\ell, (c,i)$ at step 1 of $\mathcal{R}'$ such that $x \neq
x'$, then $i$
did not ignore some message from $j$ during session $(c,i)$,  $(j, \ell, c, x)$
will never be removed from $\A_i$, and eventually $j$ will be added to
$D_i$
by line 2 in the $\DMM_i$ protocol.
Hence, $j$ is shunned by $i$ starting in session $(c,i)$.
Thus, if no process is shunned by $i$ for the first time in $(c,i)$, it
must be the case that, for each process $\ell
\in \hat{M}$, all the values broadcast by processes in $\hat{L}_\ell$
agree with $f_\ell$.  Since
there will eventually be at least $t+1$ values broadcast from
processes in $\hat{L}_\ell$, all
nonfaulty processes will interpolate $f_\ell$ for all $\ell \in
\hat{M}$, and subsequently will interpolate $f$ and the secret $s$.

\textbf{Weak and Moderated Binding}.
If the dealer is nonfaulty, it follows from Validity that Weak Binding
holds, taking $r=s$.
So suppose that the dealer $i$ is faulty.
If there is
a faulty process $j$ such that $(j, c,i,
x) \in \D_\ell$ for a nonfaulty process $\ell$ and $j$ sends a message
$\ell,x', (c,i)$
in step 1 of $\mathcal{R}'$
such that $x \neq
x'$. In this case $\ell$ did not ignore a message from $j$ during session
$(c,i)$, $(j, c, i, x)$ will never be removed from $\D_\ell$, and
eventually $j$ will be added to $D_\ell$
by line 3 in the $\DMM_\ell$ protocol.
Hence, $j$ is shunned by $\ell$ starting in session $(c,i)$,
so weak and moderated binding holds.
On the other hand,
if,
for each nonfaulty process $\ell
\in \hat{M}$, all the values broadcast by processes in $\hat{L}_\ell$
are what they were expected to be
then, at the time that the first nonfaulty process
completes protocol $\mathcal{S}'$,
the set $\hat{M}$ is fixed.
Let $H \subseteq \hat{M}$ be the set of nonfaulty processes in
$\hat{M}$. For each $\ell \in H$, the value $\hat{f}_\ell(0)$ is also
fixed. If there exists a degree-$t$ polynomial $h$ that interpolates the
points in
$\{(\ell,\hat{f}_\ell(0)) \mid \ell \in \hat{M}\}$, then let
$r=h(0)$; otherwise, let $r=\bot$.
We claim that each nonfaulty process will output either $r$ or $\bot$ at
the reconstruct phase. This is true since all nonfaulty processes will
interpolate $\hat{f}_\ell$ for all $\ell \in \hat{M}$ correctly. Since
$|H|\geq t+1$, the values $\{(\ell,\hat{f}_\ell(0)) \mid \ell \in
\hat{M}\}$ determine a polynomial $h$. If all remaining values
$\bar{f}_\ell(0)$ obtained from the polynomials $\bar{f}_\ell$ for $\ell
\in \hat{M} \setminus H$ agree with $h$, then $r$ is
output;  otherwise,  $\bot$ is output.

It easily follows from step 5 of
$\mathcal{S}'$ that
if the moderator is nonfaulty, then
the values $\{(\ell,\hat{f}_\ell(0)) \mid \ell \in
\hat{M}\}$ can be interpolated only by a polynomial $h$ such that $h(0)$ is the moderator's value $s'$; that is, $r = s'$.

\textbf{Moderated Hiding}.   If the dealer and moderator are nonfaulty
then, as long as no nonfaulty process has invoked protocol $\mathcal{R}$,
the combined view of any $t$ faulty processes is distributed
independently of the value of the shared secret, $s$.
This follows since the dealer uses random degree-$t$ polynomials,
so no set of size $t$ learns any information.
\end{proof}

As promised, we now show that it is possible that two nonfaulty
processes will complete an invocation of \mwdavss with different values
without detecting a new faulty process.
\begin{example}\label{xam:shunning}
{\rm Let $n=4$ and $t=1$. Consider an invocation of the \mwdavss protocol with
processes $1$, $2$, $3$, and $4$, where $2$ is the dealer and $1$ is the
moderator. Suppose that, in the share protocol
$\mathcal{S}'$, process $4$ is delayed.
Hence, processes $1$, $2$, and $3$ hear only from each other before
completing the  share protocol.
Thus, $L_1 = L_2 = L_3 = M = \{1,2,3\}$.
Now suppose that in
the reconstruct protocol $\mathcal{R}'$, process $3$ hears the values
sent by $2$ according to line 1 of $\mathcal{R}'$ before hearing from
$1$ or $4$.  Since it clearly hears from itself as well, $K_{3,1}$,
$K_{3,2}$, and $K_{3,3}$ will each have two points---one from $2$ and
one from $3$.  Since $t+1 = 2$ in this case, it follows from step 3 that
$3$ will then find the unique degree 1 polynomials
$\hat{f}_1$, $\hat{f}_2$, and $\hat{f}_3$ that
interpolate the points in $K_{3,1}$, $K_{3,2}$, and $K_{3,3}$, respectively.
If $\hat{f}_1(0)$, $\hat{f}_2(0)$, and $\hat{f}_3(0)$ are collinear, and
$\bar{f}$ is the polynomial that interpolates them, then $3$ outputs
$\bar{f}(0)$.  If $2$ is faulty, then by choosing
the values it sends appropriately, $2$ can make $\bar{f}(0)$ an
arbitrary element of $F$.  Now if $1$ hears from $3$ before
hearing from $2$ or $4$,
$1$ will also output a value, which may be different from $3$'s.

Of course, to get $3$ to output a value different from $1$'s,
$2$ must send a value $\hat{f}^2_1$ that is different from the one that
$1$ expects to
hear.  Once $1$ gets this value, it will realize that $2$ is
faulty, and add $2$ to its set $D_1$.  However, this may happen after
both $2$ and $3$ have completed the invocation of \mwdavss.
Notice that this argument relies on the fact that processes use RB
to send their values.  } \qed
\end{example}

\section{Implementing \davss}\label{sec:FD-AVSS}
In this section, we show how to implement \davss, and then prove that our
implementation satisfies the \davss properties.
The difficulties of doing this are illustrated by Example~\ref{xam:shunning}:
it is possible that two nonfaulty processes output different values in
an invocation $(c,i)$ of the
\mwdavss protocol. Of course, by the Weak Binding property, this can
happen only if a new faulty process is eventually detected (and is
shunned in all invocations that follow $(c,i)$).
Nevertheless, this detection can come after all processes have
completed $(c,i)$.  Thus, we must show that the inconsistency cannot
cause problems.

\subsubsection*{Share protocol $\mathcal{S}$:}
\begin{enumerate}
\item If a dealer $i$ wants to invoke $\mathcal{S}$ with a
secret $s$, it first updates $c$ to $c+1$,
initializes sets of processes $G_{(c,i)}$ and $G_{j,(c,i)}$, $j \ne i$,
to $\emptyset$
and
chooses
a random degree-$t$ bivariate polynomial
$f(x,y)$ over the field $F$ such that $f(0,0)=s$%
\footnote{Specifically, since a bivariate polynomial of degree $t$ has
the form $\sum_{i=0}^t \sum_{j=0}^t a_{ij} x^i y^j$, we simply set $a_{00} = s$
and choose the remaining coefficients at random from $F$.  Of course,
the same ideas apply to choosing a  random univariate polynomial $f$
such that $f(0) = s$.}
Let $g_j(y) = f(j,y)$ and let $h_j(x) = f(x,j)$, for $j = 1, \ldots, n$.
Dealer $i$ sends each process $j$
the message $g_j(1), \ldots, g_j(t+1), h_j(1), \ldots, h_j(t+1), (c,i)$
(so $j$ can reconstruct $g_j$ and $h_j$).

\item If a process $j$ receives $g_j$ and $h_j$ from dealer $i$ for a
session $(c,i)$, then for each process $\ell \ne j$, process $j$
participates in four invocations of \mwdavss protocol $\mathcal{S}'$:

    \beginsmall{enumerate}
    \item as a dealer with secret $f(\ell,j)$ and moderator
    $\ell$
(who should also have value $f(\ell,j)$  if $i$ and $\ell$ are
    nonfaulty);
    \item as a dealer with secret $f(j,\ell)$ and moderator
    $\ell$ (who should also have value $f(j,\ell)$ if $i$ and
    $\ell$ are nonfaulty);
    \item as a moderator with secret $f(\ell,j)$ and dealer
    $\ell$ (who should also have value $f(\ell,j)$  if $i$ and
    $\ell$ are nonfaulty); and
    \item as a moderator with secret $f(j,\ell)$ and dealer
    $\ell$ (who should also have value $f(j,\ell)$  if $i$ and
    $\ell$ are
    nonfaulty).
    \endsmall{enumerate}
\item The dealer $i$ adds $j$ to the set $G_{\ell,(c,i)}$ and $\ell$ to the
set $G_{j,(c,i)}$ if
the dealer completes all four invocations of
the share part of
\mwdavss $\mathcal{S}'$ with
$j$ and $\ell$ playing the roles of dealer and moderator.

\item The dealer $i$ adds $j$ to the set $G_{(c,i)}$ if $|G_{j,(c,i)}|\geq n-t$.

\item If $|G_{(c,i)}|\geq n-t$, the dealer sends $G_{(c,i)}, \{G_{j,(c,i)}  \mid j \in
G\}, (c,i)$ using RB.

\item When process $\ell$ receives $\hat{G}, \{\hat{G}_j  \mid j \in G\}, (c,i)$ from the dealer and completes all four $\mathcal{S}'$ protocols
for each pair $j,\ell$ such that $j
\in \hat{G}$ and $\ell \in \hat{G}_j$, then it completes this invocation of $\mathcal{S}$.
\end{enumerate}

\subsubsection*{Reconstruct protocol $\mathcal{R}$:}

\begin{enumerate}
\item Each process
$j$ initializes the set $I_{j,(c,i)}$ to $\emptyset$
and invokes the reconstruct protocol $\mathcal{R}'$
for
each of the four invocations of \mwdavss for each
pair $(k,\ell)$ such that $k \in G_{(c,i)}$ and
$\ell \in G_{k,(c,i)}$.
After the four reconstruct protocols associated with $k$ and $\ell$ are
complete,
$j$ sets $r^j_{x,k,\ell,(c,i)}$ to
    the reconstructed
    output value for the entry $f(k,\ell)$ where $x$ was the dealer in the
\mwdavss protocol (so that $x$ is either $k$ or $\ell$).

\item For each $k \in G$, process $j$ adds $k$ to $I_{j,(c,i)}$ if
\begin{itemize}
    \item there exists $\ell \in G_k$ such that $r_{kk\ell}$ or $r_{k
    \ell k}$ are $\bot$; or

    \item there do not exist degree-$t$ polynomials that interpolate
    $\{(\ell, r^j_{kk\ell}) : \ell \in G_k \}$ or
    $\{(\ell, r^j_{k\ell k}) : \ell \in G_k\}$.
\end{itemize}
Intuitively, $I_{j,(c,i)}$ consists of those processes that $j$
ignores in invocation $(c,i)$.

\item For each $k \in G \setminus I_j$, process $j$ computes the
degree-$t$ polynomials $g_k$ and $h_k$
 that interpolate  $\{(\ell,r^j_{kk\ell}) :
 \ell \in G_k \}$ and $\{(\ell, r^j_{k\ell k}) : \ell \in
G_k\}$.
If there exist $k,\ell \in G \setminus I_j$ such that $h_k(\ell)
  \neq g_\ell(k)$,
then
$j$ outputs $\bot$.
Otherwise,
if there is  a
unique degree-$t$ bivariate polynomial $\bar{f}$
  such that for all $k,\ell \in G\setminus I_j$,
  $\bar{f}(k,\ell)=g_k(\ell)=h_\ell(k)$,
then $j$ outputs $\bar{f}(0,0)$; otherwise, $j$ outputs $\bot$.
\end{enumerate}

This completes the description of the \davss protocol.

\begin{lemma}\label{lem:d-avss}
The \davss protocol satisfies the \davss properties.
\end{lemma}
\begin{proof}
For any \davss session $(c,i)$, if $k,j$ are nonfaulty processes, then
all messages sent from $k$ to $j$ will eventually not be ignored. This
is true since, if $(c',i') \rightarrow_j (c,i)$, then $j$
completed all $\mathcal{R}'$ invocations associated with $(c,i)$. From
the way we use \mwdavss in $\mathcal{R}$, all processes will also invoke
all $\mathcal{R}'$
sessions
 associated with $(c,i)$. Hence from the Termination
property of \mwdavss and \lemmaref{lem:DMM}, it follows that all
messages that $j$ expects $k$ to send in session $(c',i')$ will eventually be
received.
We now go through \davss properties in turn.

\textbf{Validity of Termination}. If the dealer is nonfaulty, then for
any two nonfaulty processes $k$ and $\ell$, eventually all four
invocations of
$\mathcal{S'}$ will complete. So eventually the set $G_\ell$
will be of size at least $n-t$ for each nonfaulty $\ell$, the set $G$
will eventually contain at least $n-t$ elements, and all four
$\mathcal{S}'$ invocations for each $j \in G$ and $\ell \in G_j$ will
complete. By the properties of RB, all processes
will eventually receive the sets $G$ and $\{G_j : j \in G\}$ and, by the
Termination property of \mwdavss, for each  $j \in G$ and $\ell \in G_j$,
all processes will eventually complete all four invocations of $\mathcal{S'}$.
Hence, all nonfaulty processes will complete protocol
$\mathcal{S}$.

\textbf{Termination}. If a nonfaulty process completes protocol
$\mathcal{S}$,
then it
follows from the Termination property of the \mwdavss protocol and the
Reliable Broadcast properties
that all nonfaulty processes complete $\mathcal{S}$.
The fact that they all  complete
$\mathcal{R}$ follows from the Termination property of the \mwdavss protocol.

\textbf{Validity}. Suppose that the dealer $i$ is nonfaulty in an
invocation of $\mathcal{S}$ with session id $(c,i)$.  There are
two cases. If a faulty process $j$ is first shunned by a nonfaulty
process $\ell$ in some \mwdavss invocation with session $(c',i')$ that is
part of the \davss invocation with session id $(c,i)$, then,
because $\ell$ started $(c,i)$ before starting $(c',i')$ and $\ell$ completes
$(c',i')$ before completing $(c,i)$,
$j$ is
also
first shunned by $\ell$ starting in session $(c,i)$ of \davss.
On the other hand, if no faulty process is shunned starting in session
$(c,i)$, then all invocations of \mwdavss
must satisfy the first clause of the Validity and Weak and Moderated
Binding properties.
It follows from (the first clause of) the Validity property that
if $k \in G_{(c,i)}$ is nonfaulty, then for all $\ell
\in G_k$, it must be
the case that $r^j_{kk\ell} = f(k,\ell)$ and $r^j_{k\ell k} =
f(\ell,k)$ (since $k$ acts as the dealer in computing these values, $\ell$ acts as the moderator and
the values themselves are correct, since they were received from $i$).
Thus, it follows that $k \notin I_{j,(c,i)}$.
Similarly, it follows from (the first clause of) the Weak and Moderated
Binding property that, for all $k \in  G$ and $\ell \in G_k$, if
either $\ell$
or $k$ are nonfaulty, then it must be
the case that $r^j_{\ell\ell k}$ and $r^j_{k\ell k}$ are each either
$f(\ell, k)$ or $\bot$, and that
$r^j_{\ell k \ell}$ and $r^j_{kk\ell}$ are each either $f(k, \ell)$ or
$\bot$.  (Here we use the
fact that the nonfaulty process
---either $k$ or $\ell$---is acting as
either dealer or moderator in the
invocations of \mwdavss during which these values are computed.)
Thus, even if $\ell$ is faulty,  if $\ell
\notin
I_j$, then we must have
$h_k(\ell) = g_\ell(k)$
for all nonfaulty $k \in G$.
It follows  that,
in step 3 of $\mathcal{R}$, $j$
correctly reconstructs
$h_\ell$ and $g_\ell$
for all $\ell  \in G \setminus I_j$.
Thus, the polynomial $\bar{f}$ computed by $j$ will be $f$, and $j$ will
output $f(0,0)$.

\textbf{Binding}.
If the dealer is nonfaulty, it follows from Validity that Binding
holds, taking $r=s$.
If the dealer is faulty, there are again two cases.
If a faulty process $j$ is shunned by a nonfaulty
process $\ell$ in some \mwdavss invocation with session $(c',i)$ that is
part of the \davss invocation session $(c,i)$,
then, as argued in the proof of Validity, $j$ is also first shunned by
$\ell$ in invocation $(c,i)$.
On the other hand, if no
faulty process is shunned starting in session
$(c,i)$, then all invocations of \mwdavss must satisfy the first clause
of the Validity and Moderated Weak Binding properties.
Consider the time that the first nonfaulty process
completes protocol $\mathcal{S}$. At this time, the set $G$ is fixed.
Let $H$ be the set of nonfaulty processes in $G$.  Since $|G| \ge n-t$,
we must have that $|H|\ge t+1$.
If there is a unique degree-$t$ bivariate polynomial $\bar{f}(x,y)$
induced by the entries $r_{jj\ell}, r_{j\ell j}$ for all $j \in H$ and
$\ell \in G_j$, then set $r=\bar{f}(0,0)$;
otherwise, set $r=\bot$.

We claim that each nonfaulty process will output $r$ at the reconstruct
phase.  As in the proof of the Validity property for \davss,
it follows from (the first clause of) the Validity property for
\mwdavss that if $k \in G_{(c,i)}$ is nonfaulty, then
for all nonfaulty $\ell \in G_k$, we have
that $r^j_{kk\ell} = \hat{g}_k(\ell)$  and $r^j_{k\ell k} =
\hat{h}_k(\ell)$, where $\hat{g}_k$ and $\hat{h}_k$ are the polynomials sent
by $i$ to $k$.
Thus, $k \notin I_{j,(c,i)}$.
Hence, if $r=\bot$, then all nonfaulty processes will output $\bot$.
Moreover,
if $r \neq \bot$, then,
as in the proof of Validity for \davss,  by the Weak and
Moderated Binding property,
and from the fact that $|H|\geq t+1$,
for all $\ell \in G
\setminus I_j
$, it must be
the case that $g_\ell$ and $h_\ell$ agree with $\bar{f}$.
Therefore $j$ will interpolate $\bar{f}$ and output $r$.

\textbf{Hiding}.   If the dealer is nonfaulty and no
nonfaulty process has invoked protocol $\mathcal{R}$,
then the combined
view of any $t$ processes is distributed independently of the
dealer's value $s$, because every polynomial $h_j$ and $g_j$ has degree
$t$, and no process learns more than $t$ values of these polynomials.
\end{proof}

This completes the construction of the \davss protocol.  We now briefly sketch how, using ideas from Canetti and Rabin
\citeyear{CR93}, we can use \davss to construct the required
asynchronous Byzantine agreement protocol.

\section{From \davss to Byzantine\\ Agreement}\label{sec:restofprotocol}

Once we have \davss, we can get an almost-surely terminating polynomial
protocol for Byzantine agreement with optimal resilience, following the
ideas outlined in Canetti's \citeyear{Can-thesis} thesis.  We proceed
in two steps.
The first step is to get a common coin.  Canetti and Rabin
showed that, given $\epsilon > 0$, an AVSS protocol that terminates with
probability $1-\epsilon$ could be used to construct a protocol
CC that gives a common coin and terminates with probability $1-\epsilon$.
We use \davss to get a
\emph{shunning Common Coin} (SCC) protocol.
\begin{definition}[SCC]
Let $\pi$ be a protocol where each party has a random input and a
binary output. As in \davss, we tag each invocation of $\pi$ with a
unique session identifier $c$.
We say that $\pi$ is a \emph{shunning, terminating, $t$-resilient
Common Coin protocol}
(SCC protocol)
if the  following properties,
called the \emph{SCC properties},
hold
(in runs with at most $t$ faulty processes
in some session tagged $c$
):
\begin{enumerate}
\item \textbf{Termination}. All nonfaulty processes terminate.

\item \textbf{Correctness}. For every invocation either
    \begin{itemize}
        \item for each $\sigma\in\{0,1\}$, with probability at least $1/4$, all nonfaulty processes output $\sigma$; or

        \item there exists a nonfaulty process $i$ and a faulty process $j$ such that $j$ is shunned by $i$ starting in session $c$.
    \end{itemize}
\end{enumerate}	
\end{definition}

\begin{lemma}
For $n>3t$ there exists a shunning, terminating, $t$-resilient Common Coin protocol.
\end{lemma}

\begin{proof}
The protocol to implement SCC is exactly the protocol in Figure 5--9
in \cite{Can-thesis}, except that we replace the AVSS protocol
with our \davss. The proof
that this protocol satisfies the SCC properties
follows from
Lemmas 5.27--5.31 in \cite{Can-thesis}, together with the
observation that if a process is shunned starting at a \davss invocation
whose reconstruct protocol competes before the SCC protocol invocation
completes, then this process is shunned starting at this SCC protocol
invocation.
\begin{comment}
following observation.

%
%
%
%
%
%
%
%
%
%
%
%
%
%
%
%
%
%
%
%
%
%
%
 If a \davss share invocation $(c',i')$ is an attached secret of a
process $j$ (see line 2 of figure 5--9 for the definition of attach) for
some invocation $(c,i)$ of protocol SCC, and $j$ belongs to some set
%
%
%
$\mathcal{G}_{j'}$ (see figure 5--9 for the use of $\mathcal{G}$) then the
invocation $(c',i')$ of \davss must begin after the invocation $(c,i)$
of FD, and $(c',i')$ must end before $(c,i)$ does.
Hence, if some faulty process $j_1$ is shunned by some nonfaulty process
$j_2$ in $(c,',i')$, then $j_1$ is also shunned by $j_2$ in $(c,i)$.
%
%
%
%
On the other hand,
%
%
%
if no such \davss invocation $c'$ is shunned during invocation $c$,
then the arguments of
Canetti apply.
\end{comment}
\end{proof}

The second step is to use the common coin protocol to get the Byzantine
agreement
protocol.  Canetti and Rabin use their
common coin protocol CC that terminates with probability $1-\epsilon$ to
get a Byzantine agreement protocol that
terminates with probability $1-\epsilon$.  We replace the use of
CC by SCC to get an almost-surely terminating protocol.
The key observation is that in the protocol of Figure 5-11 in
\cite{Can-thesis}, if a nonfaulty process $j$
participates in rounds $r$ and $r'$ (and hence, in our setting, it
participles in the SCC protocol with session identifiers $r$ and
$r'$),
and $r < r'$,
then it must be the case that
$r \rightarrow_j r'$.
Therefore, there can be at most $t(n-t)=O(n^2)$ rounds $r$ such that a
nonfaulty process $i$  shuns a faulty process $j$ starting in round $r$.
Hence, there are
at most $O(n^2)$ rounds where the SCC protocol
does not succeed. In all the remaining rounds, the first clause of the SCC
Correctness property holds,
so we essentially have a common coin that is sufficiently strong for
Byzantine agreement.
It therefore follows from Lemma 5.38 and
5.29 of \cite{Can-thesis} that
the expected running time of the protocol is  $O(n^2)$.
Thus we have the following result.

\begin{theorem}[Byzantine Agreement]
There is an
almost-surly terminating, polynomial protocol for
asynchronous Byzantine agreement protocol with optimal resilience.
\end{theorem}

\section{Conclusions}

We have shown how to use \davss to give a protocol for asynchronous
Byzantine agreement that has optimal resilience, almost-surely
terminates, and is polynomially efficient.
Our \davss protocol has implications for asynchronous Secure Multiparty
Computation  (ASMPC) of certain functionalities. In the full paper we
define a family of functionalities for which the use of \davss gives a
protocol for
ASMPC that has optimal resilience, terminates almost surely, and has
perfect security (the ideal and real worlds are statistically
indistinguishable).
Perhaps the major open question remaining is whether there exists
an asynchronous Byzantine agreement protocol with optimal resilience and
constant expected running time.
\appendix
\section{Basic tools}
\subsection{Weak Reliable Broadcast}
A protocol $\mathcal{B}$ with a distinguished dealer holding input
$s$ is a $t$-tolerant \emph{Weak Reliable Broadcast
protocol} if the following holds for every execution with at most $t$ faulty processes:
\begin{enumerate}
\item \textbf{Weak termination}. If the dealer is nonfaulty, then every
nonfaulty
process will eventually complete protocol $\mathcal{B}$.
\item \textbf{Correctness}.
\begin{enumerate}
\item
if a nonfaulty process completes protocol $\mathcal{B}$, then once the
first nonfaulty process completes the protocol
there is a value $r$ such that
each nonfaulty process that completes protocol $\mathcal{B}$ accepts
$r$;
\item if the dealer is nonfaulty, then
each nonfaulty process that completes protocol $\mathcal{B}$ accepts $s$.\end{enumerate}
\end{enumerate}
\begin{lemma}
For $n>3t$ there exists a
$t$-tolerant \emph{Weak Reliable Broadcast protocol}.
\end{lemma}
\begin{proof}
This protocol, which we call WRB, is essentially Dolev's \citeyear{D82}
\emph{crusader agreement}.
It uses two types of messages; \emph{type 1 messages} have
the form $(r,1)$ and \emph{type 2 messages} have the form $(r,2)$.
WRB proceeds as follows:
\begin{enumerate}
\item The dealer sends $(s,1)$ to all processes.
\item  If process $i$ receives a type 1 message $(r,1)$ from the
dealer and it never sent a type 2 message, then process $i$ sends
$(r,2)$ to all processes.
\item If process $i$ receives $n-t$  distinct type 2 messages
$(r,2)$, all with value $r$,
then it accepts the value $r$.
\end{enumerate}

If the dealer is nonfaulty, then
it is immediate that every nonfaulty process will send $(s,2)$, and thus
will accept $s$ (since there are at most $t$ faulty processes, by
assumption).
Moreover, if the dealer is nonfaulty,
the only type 2 message sent by a nonfaulty process is $(s,2)$, so no
nonfaulty process will receive more than $t$ type 2 messages $(r,2)$ with $r \ne s$.\footnote{We assume that,
as in VSS,
if there are multiple invocations of WRB,
messages are tagged with an invocation number,
so that messages from old invocations will not be confused with messages
from the current invocation.}

To see that WRB satisfies the correctness property,
suppose, by way of contradiction, that one nonfaulty process $i$ accepts $r$
and another nonfaulty process $j$ accepts $r'$, with $r \ne r'$.  Then
$i$ must have received $n-t$ type 2 messages with value $r$ and $j$
must have received $n-t$ type 2 messages with value $r'$.
Thus, at least $n-2t \ge t+1$ processes must have sent a type 2
message to both $i$ and $j$.
At least one of these processes must be
nonfaulty.  But the protocol ensures that a nonfaulty process will send only
one type 2 message.  This gives us the desired contradiction.
\end{proof}

\subsection{Reliable Broadcast}
A protocol $\mathcal{B}$ with a distinguished dealer holding input $s$ is a $t$-tolerant \emph{Reliable Broadcast
protocol} if the weak termination and correctness properties of the Weak
Reliable Broadcast holds, and in addition, the following property holds:
\begin{enumerate}
\item[3.]
\textbf{Termination}. For every execution with at most $t$ faulty
processes,
if some nonfaulty process completes protocol
$\mathcal{B}$ then all nonfaulty processes will eventually
complete protocol $\mathcal{B}$.
\end{enumerate}
\begin{lemma}
For $n>3t$ there exists a $t$-tolerant \emph{Reliable Broadcast (RB) protocol}.
\end{lemma}
\begin{proof}
This protocol, which we call RB, is essentially Bracha's \emph{echo
broadcast}.  It uses WRB as a subroutine.  In addition to type 1  and
type 2 messages, it uses \emph{type 3} messages, which have the
form $(r,3)$.  RB proceeds as follows:
\begin{enumerate}
\item The dealer sends $(s,1)$ to all processes using Weak Reliable
Broadcast (WRB).
\item  If process $i$ accepts  message $r$ from the dealer using WRB,
then process $i$ sends
$(r,3)$ to all processes.
\item if process $i$ receives at least $t+1$ distinct
type 3 messages with the same value $r$,
then process $i$ sends $(r,3)$ to all processes.
\item if process $i$ receives at least $n-t$  distinct
type 3 messages with the same value $r$, then it accepts the value
$r$.
\end{enumerate}

To see that RB is correct, first observe
that, from the correctness property of WRB, it follows that
it cannot be the case that two type 3 message with different values are
sent by nonfaulty processes at step 2.
Moreover, if a nonfaulty process
sends a type 3 message at step 3, it must be because it got a type 3
message from a nonfaulty process.  It easily follows that all the type 3
messages sent by nonfaulty processes at either step 2 or step 3 have the
same value.

If the dealer is nonfaulty, then it is easy to see that
all nonfaulty processes terminate and accept value $s$, as in WRB.  To
see that
termination holds for RB,
suppose that a nonfaulty process completes the
protocol.
It thus must have received $n-t$ type 3 messages with the same value
$r$.  Each other nonfaulty process
will eventually have received at least $n-2t \ge t+1$ of these messages,
and so will send a type 3 message by step 3, if it has not already
done so by step 2.  As we argued above, all the type 3 messages sent
by nonfaulty processes must have the same value.  Thus,
each nonfaulty process will end up receiving
$n-t$ type 3  messages with value $r$.

Finally, part (b) of correctness follows easily from our observation
above that all the type 3 messages sent by nonfaulty processes have the
same value $r$.
\end{proof}

\end{document}